\documentclass[conference]{IEEEtran}
\IEEEoverridecommandlockouts
\usepackage{cite}
\usepackage{amsmath,amssymb,amsfonts}
\usepackage{algorithmic}
\usepackage{graphicx}
\usepackage{textcomp}
\usepackage{xcolor}
\usepackage{subcaption}
\usepackage{geometry}
\def\BibTeX{{\rm B\kern-.05em{\sc i\kern-.025em b}\kern-.08em
    T\kern-.1667em\lower.7ex\hbox{E}\kern-.125emX}}
\begin{document}

\title{An End-Cloud Computing Enabled Surveillance Video Transmission System\\

}

\author{\IEEEauthorblockN{Dingxi Yang$^{1}$, Zhijin Qin$^{1}$, Liting Wang$^{1}$, Xiaoming Tao$^{1}$, Fang Cui$^{2}$, and Hengjiang Wang$^{1,2}$}
\IEEEauthorblockA{$^{1}$Department of Electronic Engineering, Tsinghua University, Beijing, China \\
$^{2}$China Mobile Group Device Co., Ltd. \\
E-mail: qinzhijin@tsinghua.edu.cn}
}
\maketitle

\begin{abstract}
The enormous data volume of video poses a significant burden on the network. Particularly, transferring high-definition surveillance videos to the cloud consumes a significant amount of spectrum resources. To address these issues, we propose a surveillance video transmission system enabled by end-cloud computing. Specifically, the cameras actively down-sample the original video and then a redundant frame elimination module is employed to further reduce the data volume of surveillance videos. Then we develop a key-frame assisted video super-resolution model to reconstruct the high-quality video at the cloud side. Moreover, we propose a strategy of extracting key frames from source videos for better reconstruction performance by utilizing the peak signal-to-noise ratio (PSNR) of adjacent frames to measure the propagation distance of key frame information. Simulation results show that the developed system can effectively reduce the data volume by the end-cloud collaboration and outperforms existing video super-resolution models significantly in terms of PSNR and structural similarity index (SSIM). 
\end{abstract}


\section{Introduction}
With the rapid development of video applications such as live streaming and real-time communications, Internet traffic has been dominated by video transmission \cite{b1}. In distributed surveillance applications, videos acquired by multiple cameras are delivered to the cloud server for data processing and fusion\cite{b18}. This process requires a substantial amount of spectrum resources, especially when there are plenty of surveillance cameras. Therefore, efficient video compression methods have been extensively researched and numerous video coding standards have been established such as H.264/Advanced Video Coding (AVC) \cite{H264}, H.265/ High-Efficiency Video Coding (HEVC) \cite{HEVC}, etc. These traditional video coding methods are mainly based on techniques like motion estimation, motion compensation, and entropy coding. With the development of artificial intelligence (AI) technologies, video compression methods based on deep learning have emerged. For example, Deep Video Compression (DVC) \cite{DVC} is the first end-to-end video coding method based on deep neural networks (DNNs). The Video Compression Transformer (VCT) \cite{b19} presents a transformer-based model to encode videos, outperforming existing methods including HEVC.

Inspired by the recent advances of AI, Neural Video Delivery (NAS) is a new practical video delivery framework based on DNNs and end-cloud computing \cite{b21}. In NAS, the downsampled video and content-aware DNN models trained at the cloud server are transmitted to the clients to reduce the bandwidth budget. Then the content-aware models are executed to super-resolve low-resolution (LR) videos \cite{b21} at the client devices. Specifically, a video is segmented into several chunks, and for each chunk a DNN is trained to overfit its data, ensuring high performance and user quality of experience. 
However, it's hard for cameras with limited computational resources to train models for LR chunks, making NAS unsuitable for surveillance video transmission scenarios.

Therefore, employing a unified video super-resolution (VSR) model to reconstruct high-resolution (HR) surveillance videos at the cloud has advantages over NAS. Because it greatly reduces the data volume by transmitting downsampled videos while eliminating the requirement for cameras to train models for each video chunk. In recent years, deep learning enabled VSR methods have achieved results far superior to traditional VSR methods, and related research has received widespread attention \cite{b2}. Compared to single image super-resolution, the utilization of inter-frame information significantly impacts the performance of VSR models \cite{b3}\cite{b4}. Therefore, \cite{b2} classifies VSR methods based on whether explicit alignment and specific alignment methods are used. Methods such as DUF \cite{b5} and RSDN \cite{b6} do not use explicit alignment, while TDAN \cite{b3} and EDVR\cite{b4} are alignment-based. \cite{b8} divides deep learning-based VSR models into four interrelated components, including propagation, alignment, feature aggregation, and upsampling. The propagation and alignment components describe the way to use inter-frame information, thus having the most significant impact on model performance. Based on the research of \cite{b8}, enhanced propagation and alignment components are proposed to improve the VSR performance\cite{b9}.

Both \cite{b11} and \cite{b12} propose hybrid imaging systems and each system incorporates a VSR model assisted by key frames. However, the key frames are densely distributed with a fixed interval, resulting in a substantial data volume. So it's crucial to select key frames in a sparser manner. Additionally, designing key frame selection strategies based on the video content can further optimize the utilization of key frame information.

Due to the limited quality of HR videos reconstructed from only LR videos, and considering that HR key frames can be extracted from the original video to assist VSR, we study the reconstruction of HR videos using key frames and investigate the extracting strategy of key frames. The main contributions of this paper are summarized as follows:

\begin{itemize}
    \item An end-cloud computing enabled surveillance video transmission system is developed. Specifically, the downsampled video and HR key frames acquired at the end-side cameras are transmitted to the cloud server where the inference of a unified VSR model is executed to reconstruct the original video.
    \item We propose a key frame selection strategy and a method for detecting and removing redundant frames at the end-side cameras. 
    \item We propose a key-frame assisted video super-resolution (KA-VSR) model that incorporates a key frame alignment component and an enhanced propagation mechanism. Simulations are conducted to demonstrate the effectiveness of these methods.

\end{itemize}

    \section{System model}
In this paper, the considered scenario involves end-side cameras as the sender, transmitting surveillance videos to the cloud. Nevertheless, the proposed end-cloud computing enabled system can be readily extended to other video transmission scenarios with limited spectrum resources. This section is about the proposed video transmission system model. 
\begin{figure*}[tb]
\centerline{\includegraphics[width=0.8\linewidth]{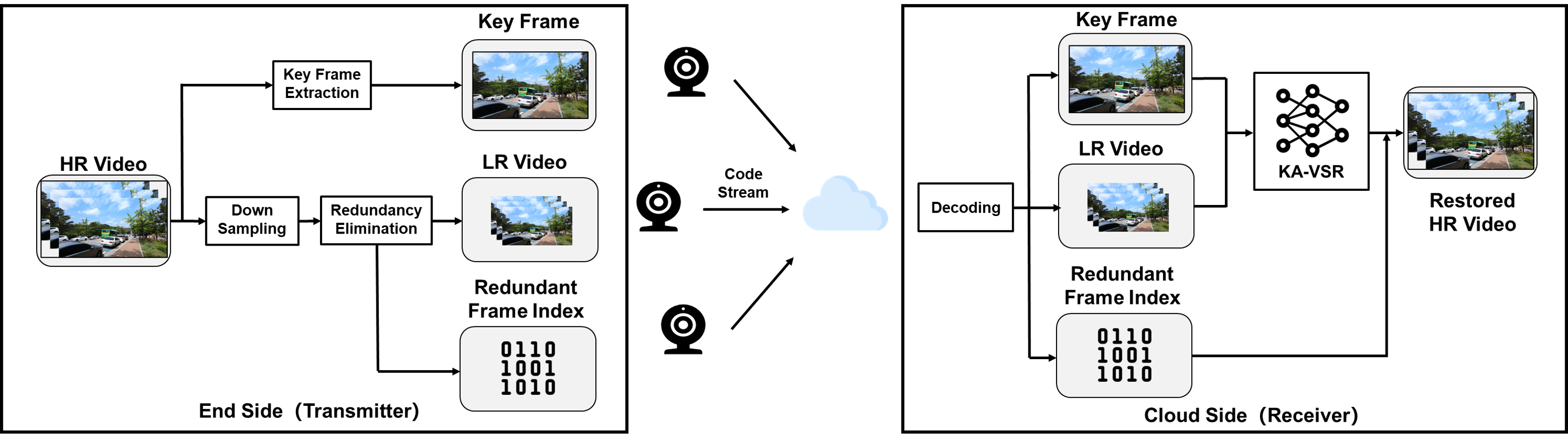}}
\caption{Framework of the proposed surveillance video transmission system enabled by end-cloud computing.}
\label{fig1}
\end{figure*}

As shown in Fig.~\ref{fig1}, the video captured by the camera is spatially downsampled at the end side. Specifically, after deconstructing the source video into a frame sequence, a 4× bicubic downsampling in the spatial dimension is performed to acquire the corresponding LR frame sequence. By downsampling, the data volume of the video is significantly reduced. The $t$-th HR frame in the frame sequence is denoted as $H_t$, and we can obtain the $t$-th LR frame $L_t$:
\[L_t(x,y)=\sum_{n=1}^{16}H_t(x_n,y_n)W(|x-x_n|)W(|y-y_n|),\tag{1}\]
where $W(\cdot)$ denotes the primary function of bicubic downsampling and $n$ denotes the 16 pixels in $H_t$ that are nearest to $L_t(x,y)$.

Secondly, key frames $\boldsymbol{K}=\{K_1, K_2,…, K_N\}$ are extracted from the HR video according to the position index $\boldsymbol{I}_k$ acquired by the key frame selection module. Since the fixed-interval selection method cannot always achieve the best performance, designing key frame selection strategies based on the video content can further optimize the utilization of key frame information.

Additionally, due to the static nature of the surveillance video scenes, numerous static frames are redundant. Therefore, it's necessary to eliminate redundant frames from the LR frame sequence. The length of the sequence is denoted as $T$ and the LR frame sequence can be represented as $\boldsymbol{LR}=\{L_1, L_2,…, L_T\}$. 
The index of redundant frame $\boldsymbol{I}_r$ can be obtained from a redundant frame detection module $g(\cdot)$:
\[\boldsymbol{I}_r=g(\boldsymbol{LR}).\tag{2}\]
Since it cannot effectively eliminate redundancy by simply removing frames that are exactly the same as their adjacent frames, we propose a redundant frame elimination module which will be detailed in Section III. The LR frame sequence after removing redundant frames is denoted as $\overline{\boldsymbol{LR}}$.

Then conventional source coding is performed on both the LR video acquired from $\overline{\boldsymbol{LR}}$ and the key frames $\boldsymbol{K}$. Additionally, the position information of the redundant frame $\boldsymbol{I}_r$ needs encoding, and the overall code stream is transmitted to the cloud server.

The cloud server decodes the received code stream to obtain the LR video $\hat{\boldsymbol{LR}}$ and HR key frames $\hat{\boldsymbol{K}}$, which are then input into a VSR model, denoted as $\mathcal{V}(\cdot)$, to acquire the restored HR frames. Then according to the received redundant frame position  $\boldsymbol{\hat{I}}_r$, the HR frames corresponding to the redundant LR frames which are removed before transmission are reconstructed by copying the non-redundant frame. Denote this copying operation as $\mathcal{C}(\cdot)$, then the complete HR reconstructed video $\hat{\boldsymbol{HR}}$ can be obtained:
\[\hat{\boldsymbol{HR}}=\mathcal{C}(\mathcal{V}(\hat{\boldsymbol{LR}},\hat{\boldsymbol{K}}),\boldsymbol{\hat{I}}_r).\tag{3}\]


\section{Proposed System Components}
The proposed end-cloud computing enabled surveillance video transmission system reduces the data volume by only transmitting the downsampled video and the key frames and utilizes a VSR model to reconstruct HR video at the cloud. In this section, we will provide a detailed description of the key frame selection strategy, the redundant frame elimination mechanism, and the KA-VSR model, which are essential components of this system.
\subsection{Key Frame Selection Strategy}
Key frames can provide a large amount of high-frequency detail information, assisting the cloud-side VSR model to enhance the reconstruction quality. Selecting key frames with a fixed interval is the simplest method and provides a relatively stable performance improvement. The fixed interval is denoted as $k$, i.e. the index of key frames is $\boldsymbol{I}_k=\{1, k+1, 2k+1, …, (N-1)k+1\}$. However, this method has a limitation in not leveraging the specific frame content to determine the optimal position of the key frames. Therefore, the fixed-interval selection method cannot always achieve the best performance.

Experiments have shown that it is an effective method to utilize the peak signal-to-noise ratio (PSNR) of adjacent frames to measure the propagation distance of key frame information. Therefore, we identify the frame where the inter-frame PSNR curve reaches its local maximum as the key frame 
 to enhance the reconstruction performance of the model. Specifically, we calculate the inter-frame PSNR between the LR frames which is denoted as $\boldsymbol{p}_{int}$, smooth this PSNR curve, and then find the position of the local maximum. Since inserting key frames too close together will result in redundant information and bring a higher cost than the performance improvement, we determine the key frame positions as sparsely as possible based on the positions of the local maximum of the PSNR curve:
\[\boldsymbol{I}_k=\mathcal{F}(\mathop{\arg\max}\limits_{t\in[1,T-1]}(s(\boldsymbol{p}_{int},w))),\tag{4}\]
where $s(\cdot)$ denotes using the Hanning window to smooth the PSNR curve with the window length $w=13$ and $\mathcal{F}(\cdot)$ denotes selecting key frames based on the positions of the local maximum in a sparse manner.

\subsection{Redundant Frame Elimination}
We design a redundant frame elimination module to further reduce the data volume and improve the efficiency of the system. There are two criteria for determining whether a frame is redundant. The first one is the MSE between this frame and the last non-redundant frame, which directly reflects the similarity between the two images. We denote the threshold for inter-frame MSE as $\tau_{int}$. However, the inter-frame MSE averages the errors over the entire image. Therefore, it cannot distinguish the cases where most of the regions are the same but the errors are concentrated in certain areas, such as a small-scale motion in fixed viewpoint surveillance videos. Therefore, the second criterion is introduced: the MSE of the motion region, denoting the threshold as $\tau_{mot}$. Particularly, the difference between these two frames is calculated and converted to a gray scale image. Then the area with pixel values exceeding the threshold $m$ in the gray scale image is recorded as a mask $M$, denoting this process as $\mathcal{W(\cdot)}$, thus obtaining the motion region:
\[M=\mathcal{W}(|L_t-L_l|,m),\tag{5}\]
where $L_t$ is the frame to be decided whether is redundant and $L_l$ is the last non-redundant frame. 
Subsequently, the MSE of the motion region is calculated by only taking the corresponding area in the two frames based on the mask, effectively detecting the small-scale motion. The entire process of detecting redundant frames is denoted as $g(\cdot)$:
\[\boldsymbol{I}_r=g(\boldsymbol{LR},\tau_{int}, \tau_{mot},m).\tag{6}\]

In the surveillance video dataset we create, redundant frames detected by this method account for approximately 20$\%$ to 30$\%$ of the total video frames.

\subsection{Video Super-Resolution Model}
\begin{figure}[tb]
\centerline{\includegraphics[width=0.9\linewidth]{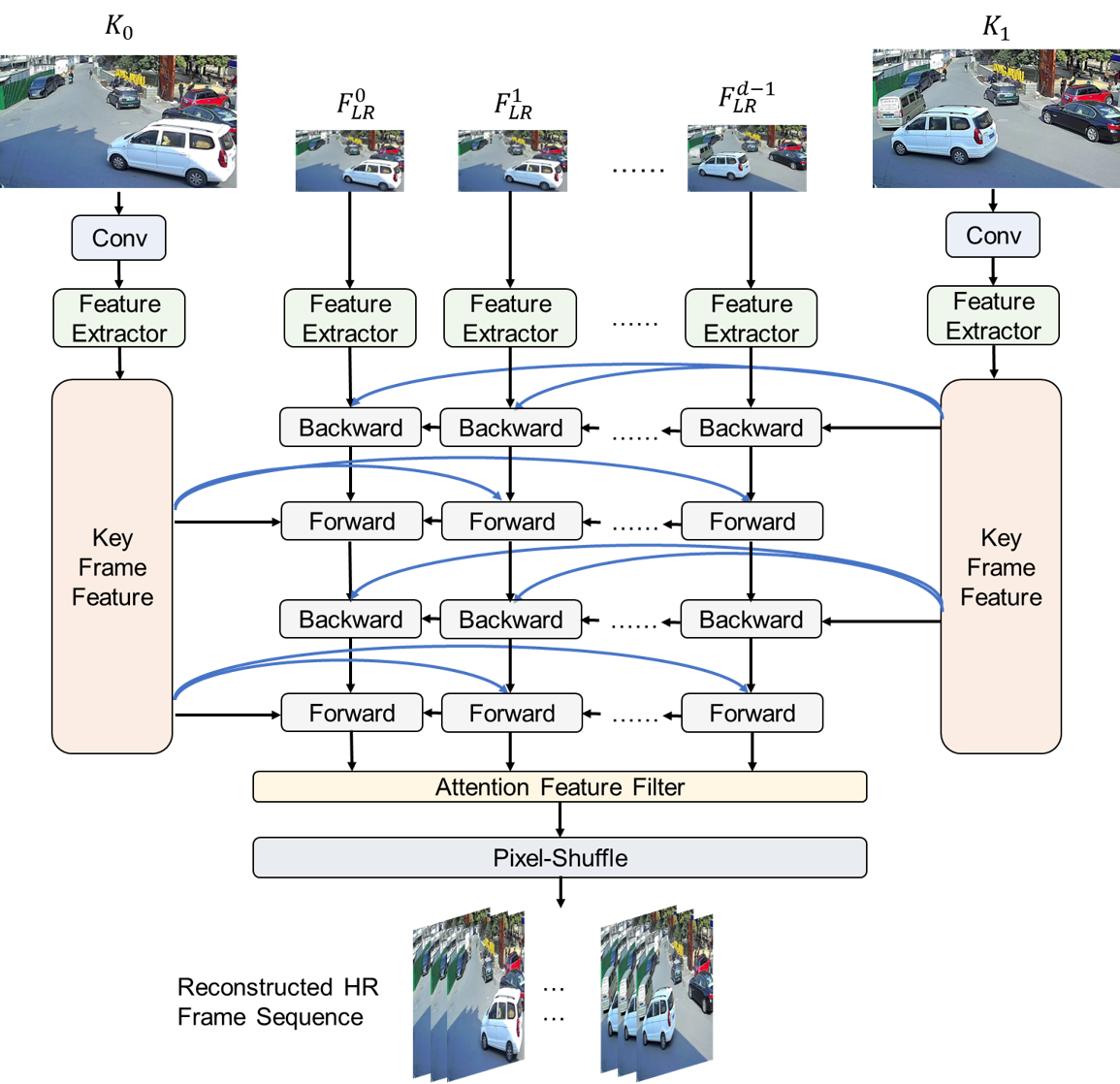}}
\caption{The overall structure of the proposed KA-VSR model.}
\label{fig3}
\end{figure}

This part provides a detailed description of the proposed KA-VSR model. The overall structure of this model is shown in Fig.~\ref{fig3}.

First, feature extraction is performed on the input LR frame sequence and the HR key frames. The basic feature extraction module consists of 5 residual blocks\cite{b9}. The $t$-th LR frame is processed by this module to extract its feature $G_t$. Since the shape of the key frames is different from the shape of the LR frames, the key frames need to be transformed into feature maps with the same width and height as the LR frames, through two consecutive 2D convolutions and Leaky-ReLU activation layers\cite{b12}. These feature maps are then processed by the aforementioned feature extraction module to obtain the key frame features $F_K^j$, where $j$ represents the index of the key frame.

Then, the extracted features are input into the Recurrent Neural Network (RNN) propagation module. Similar to BasicVSR++ \cite{b11} and NeuriCam \cite{b12}, the propagation module consists of four layers of RNN, with two layers of backward propagation and two layers of forward propagation alternately. However, unlike the second-order grid propagation in BasicVSR++, we use the first-order grid propagation combined with the propagation of key frame features. Specifically, within each layer, the first-order propagation is performed in chronological order while the key frame features are directly propagated to each LR frame. Finally, the refined features are propagated downwards layer by layer.

NeuriCam \cite{b12} also uses key frame features to assist reconstruction, where key frame feature is indirectly propagated to non-adjacent frames. However, RNN has the limitation of insufficient learning ability for long-distance contextual information. Indirect propagation will lead to ineffective utilization of key frame information. Therefore, we directly fuse the key frame features with the LR frame features to assist in reconstruction.

In the $l$-th propagation layer, there are three inputs in the propagation module of the $t$-th frame: the output feature map $F_t^{l-1}$ of the same frame position in the previous layer (where $F_t^0=G_t$, i.e., the feature extracted in the first step), the output $F_{t-1}^l$ of the previous frame in the same layer (taking forward propagation as an example), and the key frame feature $F_K^j$. Alignment is required before feature fusion, and we adopt the alignment method of flow-guided deformable convolution proposed by BasicVSR++ \cite{b11} for feature-level alignment, denoted as $\mathcal{A}(\cdot)$. Therefore, the aligned feature $\hat{F}_{t-1}^l$ from the previous frame can be represented as:
\[ \hat{F}_{t-1}^l=\mathcal{A}(F_{t-1}^l,\mathcal{S}(\hat{L}_{t-1},\hat{L}_t)),\tag{7}\]
where $\mathcal{S(\cdot)}$ represents the pretrained SPyNet\cite{b13}, which takes adjacent LR frames $\hat{L}_{t-1}$ and $\hat{L}_t$ as input to calculate the optical flow. Then the residual of the optical flow is calculated to obtain the offsets of deformable convolution (DCN)\cite{b14}. Finally, the DCN is used to align the LR frames at the feature level. As for the key frame, there is a problem of dimension mismatch. Considering that the convolutional layers that extract key frame features preserve spatial position distribution information. Therefore, the optical flow of the LR frame corresponding to the key frame can be used to guide deformable convolution to align the key frame feature $F_K^j$ with the $t$-th LR frame, improving the accuracy of key frame information propagation:
\[\hat{F}_K^j=\mathcal{A}(F_K^j,\mathcal{S}(\hat{L}_{j},\hat{L}_t)).\tag{8}\]
Actually, the offsets and modulation masks which are separately calculated for the previous frame and the key frame are concatenated, denoted as $O_i$ and $M_i$. Feature alignment is achieved using a single DCN v2 (denoted as $\mathcal{D(\cdot)}$)\cite{b15}:
\[\hat{F}_i^l=\mathcal{D}(F_{t-1}^l,F_K^j,O_t,M_t ).\tag{9}\]

Then, the feature map $\hat{F}_t^l$ and the output feature map $F_t^{l-1}$ of the same frame position in the previous layer are concatenated. After passing through a series of residual blocks, the output feature map $F_t^l$ of the $t$-th frame in the $l$-th layer is obtained. Next, the feature maps $F_t^l,(l=1,2,3,4)$ of each layer are input into the attention-based feature filter\cite{b12}, which calculates the similarity between each layer's feature map and the LR frame feature extracted at the beginning. The attention weights are then calculated using the $Softmax$ function and this process of calculating the weights is denoted as $Att(\cdot)$. The final feature map $\overline{F}_t$ is obtained by weighting:
\[\overline{F}_t=\sum_{l=1}^4Att(F_t^l,G_t )\odot F_t^l.\tag{10}\]

Finally, upsampling is performed using convolutional layers and a pixel shuffling layer\cite{b17} to obtain the reconstructed frame sequence.

\section{Simulation Results}
In this section, simulation results are presented to demonstrate the advantages of the proposed transmission system in terms of the data volume and the effectiveness of the aforementioned components.
\paragraph{Bits per pixel comparison} The advantages of the proposed video transmission system in terms of bits per pixel (bpp) are calculated as follows. We create a dataset of surveillance videos provided by China Mobile, where the videos have different resolutions. 
During transmission, the LR video is encoded using H.265, and the key frames are encoded using JPEG. The bpp of directly transmitting the HR video encoded with H.265 is compared with the proposed system. The specific data is obtained by averaging the encoding results of multiple videos or images on the surveillance video dataset, as shown in Table~\ref{tab1}.

\begin{table*}[tb]
\caption{Bpp comparison}
\begin{center}
\begin{tabular}{c|c|c|c|c|c|c|c}
\hline
\textbf{HR video} & \textbf{frame rate}& \textbf{HR video}& \textbf{LR video}& \textbf{key frame}& \textbf{key frame}&\textbf{our system}&\textbf{bpp}\\
\textbf{resolution} & \textbf{(fps)}& \textbf{bpp}& \textbf{bpp}& \textbf{interval}& \textbf{bpp}&\textbf{bpp}&\textbf{saving}\\

\hline
1280×720 &10 &\textbf{0.105} &0.011 &33 &0.064 &\textbf{0.075} &29.0\% \\
1280×720 &10 &\textbf{0.105} &0.011 &50 &0.042 &\textbf{0.053} &49.5\%\\
1280×720 &15 &\textbf{0.087} &0.008 &33 &0.064 &\textbf{0.072} &18.0\%\\
1280×720 &15 &\textbf{0.087}&0.008 &50 &0.042 &\textbf{0.050} &42.4\%\\
2304×1296 &10 &\textbf{0.066} &0.006 &33 &0.051 &\textbf{0.058} &13.2\%\\
2304×1296 &10 &\textbf{0.066} &0.006 &50 &0.034 &\textbf{0.040} &39.4\%\\
\hline
\end{tabular}
\label{tab1}
\end{center}
\end{table*}
The bpp of LR videos is approximately one-tenth of that of HR videos after a 4× downsampling. And the key frames occupy a large portion of the bpp in the proposed system, so the density of key frames is crucial. Table~\ref{tab1} shows that when the key frame interval is around 33, this system can reduce 10\%-30\% of the bpp compared to H.265. And when the key frame interval is around 50, it can reduce 35\%-50\% of the bpp. However, as the key frames become sparser, the quality of the reconstructed video will correspondingly decrease. The relevant results will be presented later in this section.

\paragraph{VSR model evaluation} For the proposed KA-VSR model, training is conducted on the REDS dataset \cite{b10} and the surveillance video dataset. Specifically, the pretrained parameters of the NeuriCam model \cite{b12} are used to initialize the parameters of KA-VSR through fuzzy matching. Then, training is performed on the REDS dataset for 100 epochs, with an initial learning rate of $1$×$10^{-4}$ for the main network and $2.5$×$10^{-5}$ for the optical flow network SPyNet \cite{b13}. The patch size used during training is 64×64. The validation clips 003, 007, 013, and 019 from the REDS dataset are used as the test set, while the other clips were used for validation. Subsequently, transfer learning is conducted on the surveillance video dataset for 100 epochs. The initial learning rate remains the same, and the patch size used during training is 80×80. 
The performance comparison is shown in Table~\ref{tab2}. Note that the first three models in the table: EDVR \cite{b4}, BasicVSR \cite{b8}, and BasicVSR++ \cite{b9} do not use key frames. Among the last three models, HIS-VSR \cite{b11} selects 1 key frame every 6 frames, while NeuriCam \cite{b12} and KA-VSR select 1 key frame every 15 frames (for performance comparison). The frames at the key frame positions restored by the models using key frames are very close to the original frames. Therefore, to calculate the average values of the PSNR and structural similarity index (SSIM) performance metrics fairly, the corresponding frames at the key frame positions are removed before calculating.
\begin{table}[tb]
\caption{Reconstruction performance comparison (PSNR/SSIM). Red indicates the best performance while blue indicates the second-best.}
\begin{center}
\begin{tabular}{c|c|c}
\hline
\textbf{} & \textbf{REDS4}& \textbf{Surveillance Video Dataset}\\
\hline
EDVR \cite{b4}& 31.09/0.8800& -  \\
BasicVSR \cite{b8}&31.42/0.8909&-\\
BasicVSR++ \cite{b9}&32.16/0.9139&24.70/0.7762\\
HIS-VSR \cite{b11}&\textcolor{blue}{32.49}/0.8841&-\\
NeuriCam \cite{b12}&31.70/\textcolor{blue}{0.9235}&\textcolor{blue}{30.79}/\textcolor{blue}{0.9466}\\
KA-VSR&\textcolor{red}{33.32}/\textcolor{red}{0.9374}&\textcolor{red}{33.54}/\textcolor{red}{0.9705}\\
\hline

\end{tabular}
\label{tab2}
\end{center}
\end{table}

\begin{figure}[t]
	\centering
	\begin{subfigure}{0.45\linewidth}
		\centering
		\includegraphics[width=1\linewidth]{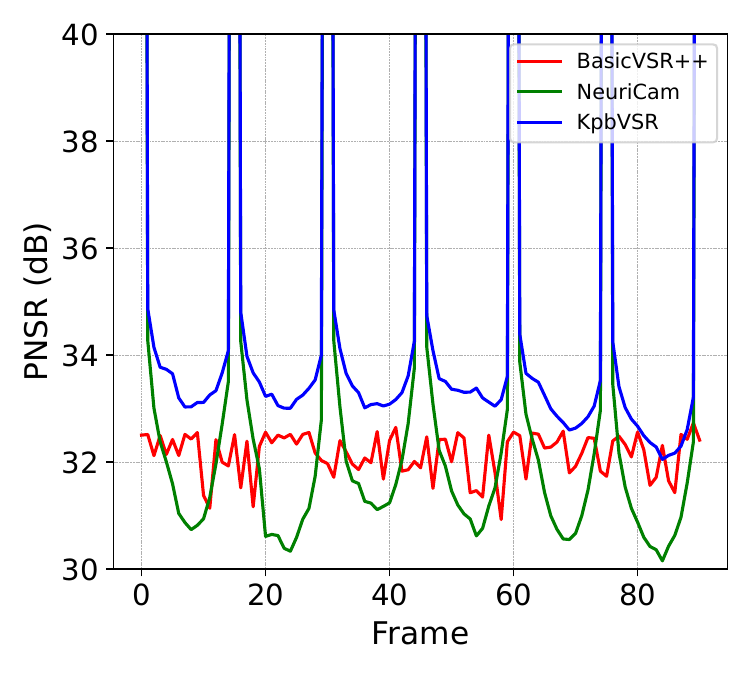}
		\caption{PSNR versus frame index}
		\label{4a}
	\end{subfigure}a
	\centering
	\begin{subfigure}{0.45\linewidth}
		\centering
		\includegraphics[width=1\linewidth]{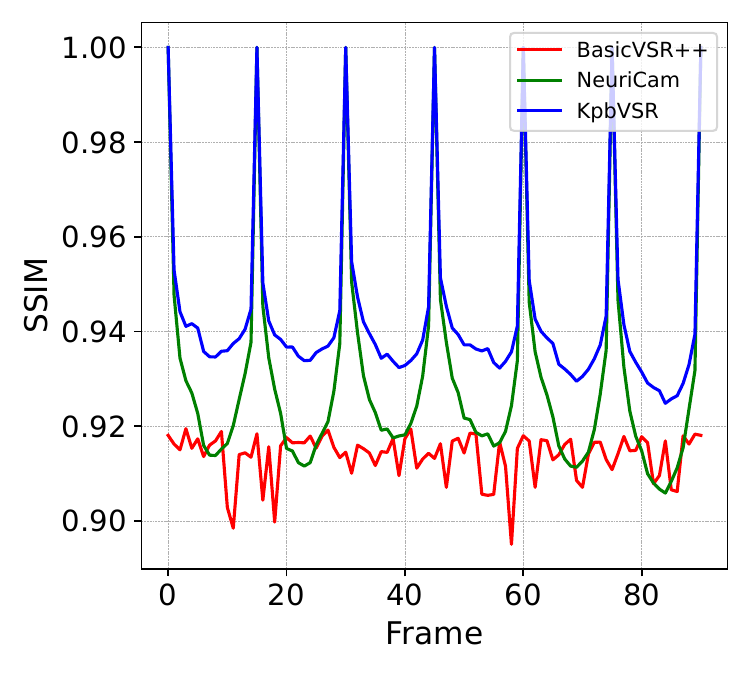}
		\caption{SSIM versus frame index}
		\label{4b}
	\end{subfigure}
	\centering
	\caption{PSNR/SSIM of frames at different positions. REDS dataset is used for evaluation and the key frame interval $k$ is set to 15.}
	\label{fig4}
\end{figure}

In general, models using key frames achieve better reconstruction performance by utilizing additional information. KA-VSR achieves the best performance in all metrics on both datasets. Fig.~\ref{fig4} further demonstrates how the key frames assist in VSR reconstruction. It shows that the reconstruction performance is significantly improved when the frame gets close to the key frames. Additionally, the proposed KA-VSR consistently outperforms the other two models in terms of average PSNR and SSIM at each position. Moreover, when the LR frames are far from the key frames, the performance degradation of the reconstructed frames in KA-VSR is significantly slower than that in NeuriCam. This demonstrates the advantage of directly propagating the aligned key frame features to other LR frames for feature fusion which is utilized in KA-VSR.

\begin{figure}[t]
	\centering
	\begin{subfigure}{0.45\linewidth}
		\centering
		\includegraphics[width=1\linewidth]{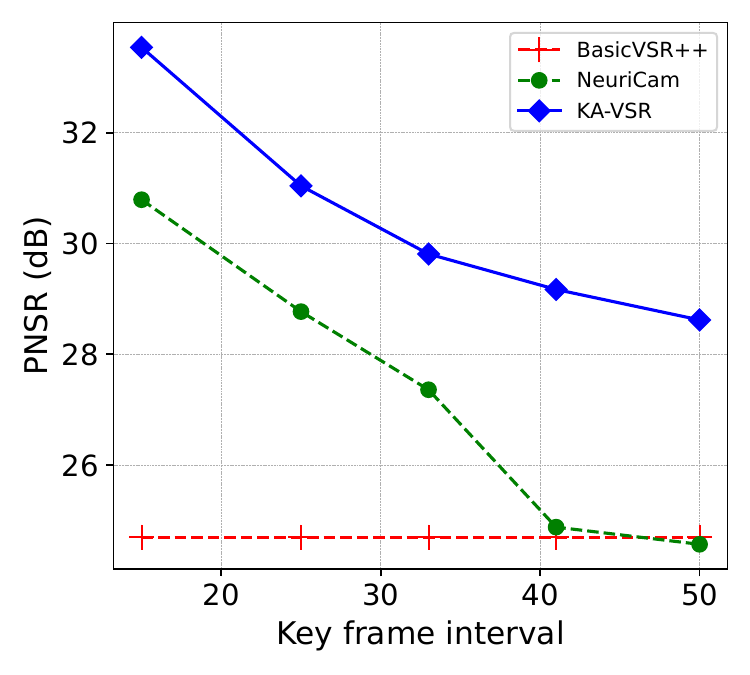}
		\caption{PSNR versus key frame interval}
		\label{5a}
	\end{subfigure}
	\centering
	\begin{subfigure}{0.45\linewidth}
		\centering
		\includegraphics[width=1\linewidth]{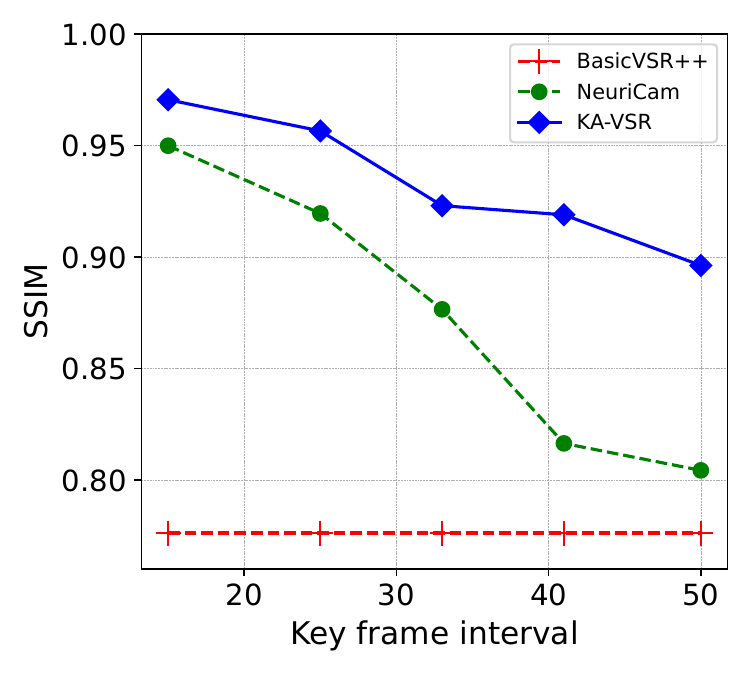}
		\caption{SSIM versus key frame interval}
		\label{5b}
	\end{subfigure}
	\centering
	\caption{Reconstruction performance with different key frame interval $k$, which is set to 15, 25, 33, 41, 50.}
	\label{fig5}
\end{figure}

\paragraph{Model performance with sparse key frames} Due to the large data volume of key frames, the proposed video transmission system won't save data volume when the key frames are dense. Therefore, experiments are conducted on the cases of different fixed key frame intervals ($k=15,25,33,41,50$). The performance comparison of the models on the surveillance dataset is shown in Fig.~\ref{fig5}. BasicVSR++ does not use key frames so it is not affected by the key frame interval. As the figure shows, KA-VSR consistently achieves the highest PSNR and SSIM at each interval. As the key frame interval increases, NeuriCam shows a rapid performance decline because its key frame features are indirectly propagated to the LR frames through RNN, making it difficult to utilize key frame information for distant frames. However, the performance degradation rate of KA-VSR is considerably slower than that of NeuriCam, once again demonstrating the effectiveness of the proposed propagation and key frame alignment methods.



\begin{figure}[t]
	\centering
	\begin{subfigure}{0.45\linewidth}
		\centering
		\includegraphics[width=1\linewidth]{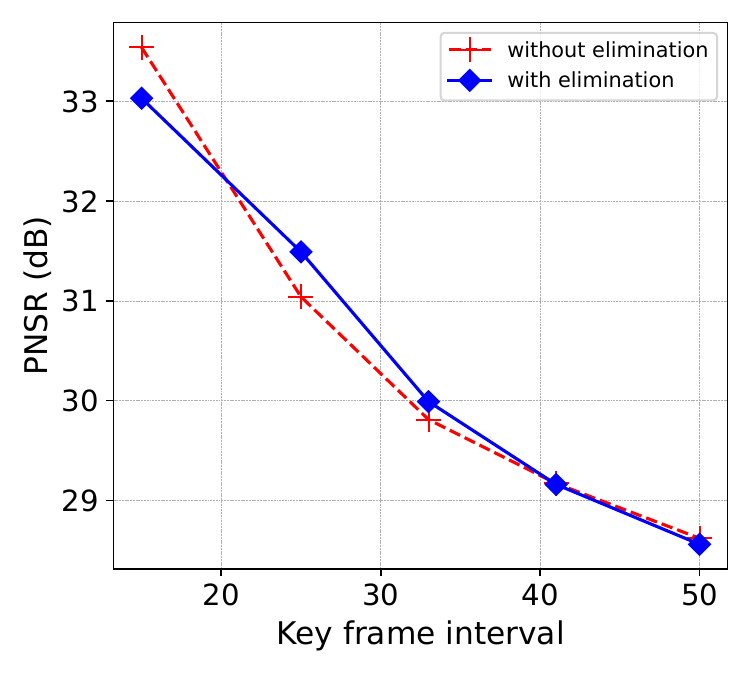}
		\caption{PSNR versus key frame interval}
		\label{6a}
	\end{subfigure}
	\centering
	\begin{subfigure}{0.45\linewidth}
		\centering
		\includegraphics[width=1\linewidth]{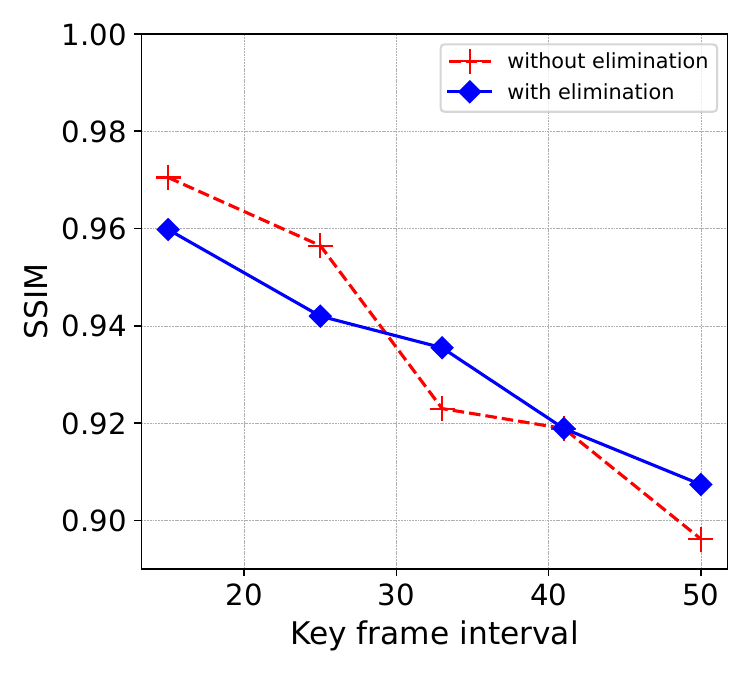}
		\caption{SSIM versus key frame interval}
		\label{6b}
	\end{subfigure}
	\centering
	\caption{Impact of removing redundant frames on reconstruction performance.}
	\label{fig6}
\end{figure}

\paragraph{Redundant frame elimination} According to experiments, the parameters of the redundant frame elimination module are chosen as follows: the inter-frame MSE threshold $\tau_{int}=0.5$, the motion region MSE threshold $\tau_{mot}=15$, and the motion mask threshold $m=2$. A 10-minute surveillance video (6000 frames in total) is randomly selected for statistics and evaluation. The number of redundant frames detected based on these thresholds is 1730, accounting for approximately 28.8\%. After eliminating the redundant frames, key frames are selected at a fixed interval. The reconstruction performances using KA-VSR with and without deleting redundant frames are compared in Fig.~\ref{fig6}. It is evident that eliminating redundant frames has a minimal impact on reconstruction performance, which demonstrates the effectiveness of the proposed module for detecting and deleting redundant frames based on the two criteria between the current frame and the last non-redundant frame. Redundant frame elimination reduces the data volume of the LR video and the number of key frames, further reducing the data volume and inference time of the model.


\paragraph{Key frame selection strategy} The experiments are setup as follows: the LR sequence length $T$ is set to 67, with the first and the last frame fixed as key frames. Another key frame needs to be selected from the other frames to improve the reconstruction performance. As for the fixed interval key frame selection method, this key frame is the $34^{th}$ frame. Following the aforementioned method, the experiment is conducted by selecting the local maximum point of inter-frame PSNR as the key frame index. The results are shown in Table~\ref{tab5}. This method improves the reconstruction performance in most of the test clips, and the average PSNR of the entire test set is improved by 0.2 dB.

\begin{table}[tb]
\caption{Impact of key frame selection strategy on PSNR (dB). Red indicates the performance improvement by strategically selecting key frames.}
\begin{center}
\begin{tabular}{c|c|c|c}
\hline
\textbf{Clips} & \textbf{Fixed interval}& \textbf{Non-fixed interval}& \textbf{Key frame index}\\
\hline
000& 31.19&\textcolor{red}{31.25}&35 \\
001&31.24&31.11&32 \\
002&26.24&26.24&34 \\
003&28.69&\textcolor{red}{29.65}&46 \\
004&29.23&28.87&47 \\
005&31.19&\textcolor{red}{31.41}&25 \\
006&30.90&\textcolor{red}{31.51}&43 \\
Average&29.81&\textcolor{red}{30.01}&-\\
\hline
\end{tabular}
\label{tab5}
\end{center}
\end{table}
\section{CONCLUSION}
In this paper, we study surveillance video transmission with limited spectrum resources. Specifically, an end-cloud computing enabled video transmission system is developed for solving the bandwidth limitation problem in surveillance transmission. The simulation results show that the transmission system can considerably reduce data volume compared to H.265 and the developed model KA-VSR outperforms all baselines significantly.

\end{document}